\documentclass{kapproc} 

\usepackage{procps}

\usepackage[dvips]{graphicx}

\upperandlowercase

\kluwerbib 

\begin{document}

\articletitle{Bars, Spiral Structure, and Secular Evolution in
Disk Galaxies}


\author{Bruce G. Elmegreen}
\affil{IBM Research Division, T.J. Watson Research Center\\
1101 Kitchawan Road, Yorktown Hts., NY 10598 USA}
\email{bge@watson.ibm.com}

\begin{abstract}
Simulations and observations of galactic bars suggest they do not
commonly evolve into bulges, although it is possible that the
earliest bars formed bulges long ago, when galaxies were smaller,
denser, and had more gas. The most highly evolved of today's bars
may become lenses over a Hubble time.  Most galaxies in the early
Universe are extremely clumpy, with $\sim10^8-10^9$ M$_\odot$ blue
clumps that resemble in color and magnitude the isolated field
objects nearby. The presence of blue and irregular bars at high
redshift suggests that some bars formed primarily in the gas phase
accompanied by giant starbursts, rather than in pure stellar disks
like most models. Secular and non-secular processes that cause
galaxies to evolve are summarized.
\end{abstract}


\section*{Introduction}
Secular evolution of galaxies includes too many processes to
review thoroughly here. Instead, a selection of processes with
contemporary interests will be discussed. They include bar
evolution in the present-day Universe and at intermediate to high
z, and disk formation at high z, where extremely clumpy structure
seems to be common. A brief summary of secular evolution effects
over a Hubble time is in Section \ref{sect-evol}. A recent review
of secular evolution leading to central mass concentrations and
pseudobulges was in Kormendy \& Kennicutt (2004).
\section{Bar Evolution}
\subsection{Bar Dissolution to Form a Bulge}
The disk and halo of a galaxy exert negative torques on a bar and
cause the bar to get stronger, longer, and slow down over time
(Weinberg 1985; Debattista \& Sellwood 1998, 2000; Athanassoula,
Lambert, \& Dehnen 2005; Athanassoula 2005). Bars also change
their shape as more orbital families get populated. They can
thicken and develop a peanut-shaped profile perpendicular to the
disk as a result of a resonance between the corotating pattern
period and the vertical oscillation period (Combes \& Sanders
1981). They can become more squared off at the ends, or boxy, as a
result of increasing populations of 4:1 orbits. Orbit families are
described in Sellwood \& Wilkinson (1993) and Athanassoula (2005).
Gas accretion to the center and subsequent star formation can lead
to a dense nucleus, which either forms an ILR or shifts the
existing ILR to a larger radius. Tightly wound starburst spirals
or rings may appear near these inner resonances and secondary bars
can form inside (e.g., Englmaier \& Shlosman 2004).

Perhaps the most startling development of the 1990's was that the
growth of the ILR radius with time depopulates bar-supporting
orbits (x1 orbits), which are the elongated orbits parallel to the
bar out to nearly the corotation radius. This depopulation can
dissolve a bar completely, forming a bulge-like concentration in
the center (Hasan \& Norman 1990; Friedli \& Benz 1993; Hasan,
Pfenniger, \& Norman 1993).  Norman, Sellwood, \& Hasan (1996) ran
2D models of bar dissolution with 50,000 star particles, no gas
and no active halo. They found that by the time 3-5\% of the total
disk mass shrank to the galaxy core, the bar was completely
dissolved. They also confirmed this result with a 3D model using
200,000 star particles and no active halo.

Shen \& Sellwood (2004) also ran 3D star models with no active
halo and found in addition that bar weakening occurs in 2 phases,
an initially fast stage where the bar-supporting orbits are
scattered, and then a slow stage where the whole potential
changes. More importantly, they pointed out that the central mass
needed to destroy a bar is much larger, by a factor of 10 or more,
than the nuclear black hole mass, concluding that bar destruction
is in fact unlikely in real galaxies. They showed further that a
dispersed central mass, as might result from gas inflow and star
formation, required an even higher fraction (e.g., $>10$\% of the
disk mass) than a point-like distribution, making the possibility
that gas accretion can eventually destroy a bar (i.e., because of
its central mass) seem even more remote.

Athanassoula, Lambert, \& Dehnen (2005) studied bars with active
halos and found that if the halo was massive, the bars survived
even with a central mass of 10\% of the disk mass.  Bars in
low-mass active halos were still destroyed, like before. The
difference is that an active halo continuously removes angular
momentum from the bar, strengthening it by populating x1 orbits.
Then the deleterious effect of the central mass is reduced. Halos
also promote spiral arm formation by the removal of disk angular
momentum (Fuchs 2004).

While bars are strengthened by a loss of angular momentum to a
disk or halo, thereby resisting dissolution at the ILR from a
central concentration, bars also weaken by the absorption of
angular momentum from gas, and thereby become more prone to
dissolution (Bournaud \& Combes 2002). Gas within the bar radius
shocks at the bar-leading dust lanes. The asymmetric pressures in
these shocks cause the gas to lose orbital angular momentum,
eventually migrating to the center. The shock is asymmetric only
if there is energy dissipation; otherwise the backward impulse the
gas feels when entering the shock is compensated by a forward
impulse it feels when leaving the shock.  From the point of view
of the stars, there is a gravitational attraction to the dust
lane, which is shifted forward of the orbital apocenter. Because
stars in bar orbits spend a longer time at apocenter than at
pericenter (on the minor axis), the time-integrated gravitational
force from the gas produces a net positive torque on the stars.
The resulting angular momentum gain causes the elongated stellar
orbits to open up into circular orbits, which have higher angular
momentum for the same energy (the same major axis length).
Circularization of orbits makes the bar ultimately dissolve.
Recent simulations of this effect are in Bournaud, Combes, \&
Semelin (2005).

Dissolution of a bar by gas torques requires a large amount of gas
in the bar region, as essentially all of the angular momentum that
was removed from the stars to make the bar in the first place has
to be returned from the gas. If the total mass fraction of gas in
the bar region is small, then the circularization of bar orbits
may be minimal. Block et al. (2002) propose that continued
accretion of gas onto the galaxy disk can cause a second or third
event of bar dissolution if a new bar forms in the mean time. In
this way, bars can dissolve but still have an approximately
constant fraction over a large portion of the Hubble time, as
observed by Sheth, et al. (2003), Elmegreen, Elmegreen, \& Hirst
(2004), Jogee et al. (2004), and Zheng, et al. (2005).

\subsection{Bars Dissolution to Form a Lens} What is the
observational evidence for bar dissolution and what happens to old
bars?  The evidence for bar circularization in more centrally
concentrated galaxies (Das et al. 2003) does not actually indicate
that bars dissolve. Dissolution may take a very long time and the
observed circularization could be only the first step.

There is good evidence for some bar dissolution, however, with the
resulting structure a lens (Sandage 1961; Freeman 1975; Kormendy
1977), or perhaps a pseudo-bulge (Kormendy \& Kennicutt 2004). A
lens is not a spheroid or part of the outer exponential disk. It
is a stellar distribution in the inner region with a rather
shallow brightness profile and a sharp cutoff at mid-disk radius.
Outside the lens, the main disk exponential continues. Kormendy
(1979) suggested that lenses are dissolved bars.  That is, S0
galaxies with lenses formerly had bars.  Of the 121
low-inclination, bright SB0-SBd type galaxies in Kormendy (1979),
54\% of early bar types, SB0-SBa, have a lens, 0\% of late bar
types SBab-SBc have a lens, and 16\% of the sample have both a bar
and a lens. The lens looks like it is coming from the bar in these
latter cases because the lens size is the same as the bar size.
The lenses in early-type non-barred galaxies are also the same
size as the inner rings in barred (non-lens) galaxies for the same
absolute magnitude galaxy; these rings are another measure of bar
size. Lens colors are generally the same as bar colors too. Some
bars even ``look'' like they are dissolving into a lens because
the bar spreads out into a circular shape at the ends (e.g., NGC
5101).  The axial ratio distribution of lenses weakly suggests
they are thick (triaxial).

The presence of lenses in about half of the early type barred
galaxies (as mentioned above) but in only a small fraction of
early type non-barred galaxies suggests further that bars dissolve
very slowly into lenses.  The most advanced stages of dissolution
are in the early type bars, which are denser and more advanced in
total evolution than late-type bars.

There are very few simulations of bar dissolution into a lens. A
pure star simulation by Debattista \& Sellwood (2000) shows
something like a stellar ring forming at the end of the bar. Stars
develop complex orbits over time, migrating between inside and
outside the bar as the lens builds up.

\section{Bar Dissolution into Bulges at Very Early Times}

We have found that barred galaxies in the early Universe are
slightly smaller than barred galaxies today, by a factor of
$\sim2$ in the exponential scale length (Elmegreen, Elmegreen, \&
Hirst 2004). Such a measurement avoids problems with cosmological
surface brightness dimming. The same small sizes are observed for
spiral galaxies in general, based on a sample of 269 spirals in
the Hubble Ultra Deep Field (Elmegreen, et al. 2005a). The
presence of spirals in these high-z disk galaxies suggests that
the disks are relatively massive compared to the halos at that
time, unlike the inner disks of today's galaxies. This suggests,
in turn, that galaxies build up from the inside out, as proposed
many years ago (e.g., Larson 1976). With such a process, the
dynamical time for evolution in a disk, which is proportional to
the inverse square root of total density, will be shorter in the
early Universe than it is today. That is, the active parts of
galaxy disks are gradually evolving toward lower and lower
densities and longer evolutionary timescales (a process related to
``downsizing'' -- e.g. Tanaka, et al. 2005).

The implication of increasing dynamical time in an aging Universe
is that bar formation and secular dissolution should have been
much faster at high redshift, in approximate proportion to the bar
size. Also, the higher gas abundance in young galaxies would have
led to a more rapid dissolution then too, by the Bournaud \&
Combes (2002) process. Thus some of the very early stages of bulge
formation inside a disk (as opposed to disk accretion around a
pre-existing bulge) could have involved evolution from a bar.
There is no direct evidence for this yet, however.

\subsection{Observing The Youngest Bars}

\begin{figure}[ht]
\vskip.2in
\centerline{\includegraphics[width=4in]{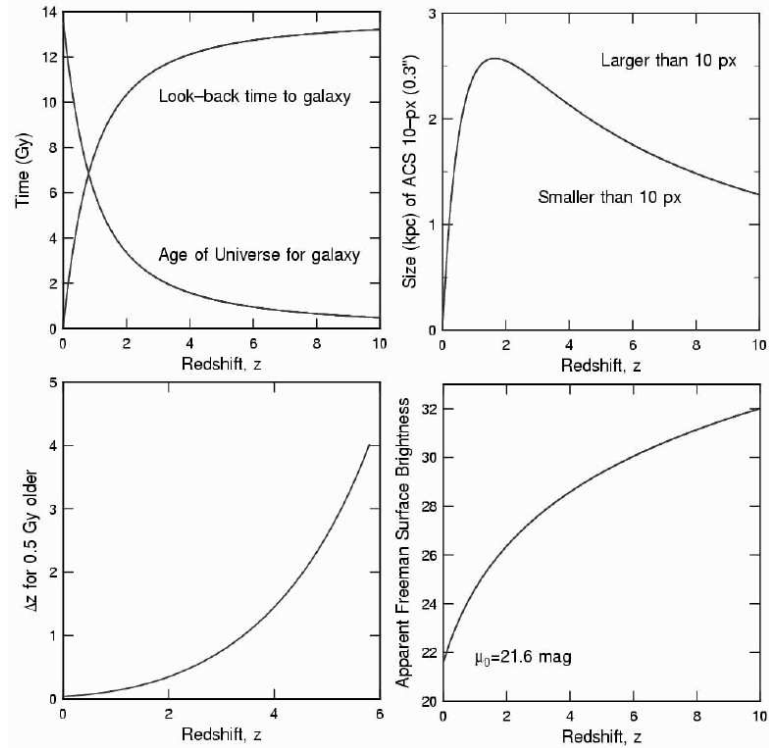}}
\caption{Four quantities from the standard $\Lambda$CDM cosmology
are plotted versus redshift, $z$: the Universe age and look back
time (top left), the physical size of a region subtending 10
pixels of the HST ACS camera (top right), the redshift range,
$\Delta z$, spanning 0.5 Gy (lower left), and the apparent surface
brightness of a region having an intrinsic surface brightness
equal to the standard Freeman value (lower right).}
\end{figure}

Observations of the youngest bars are difficult for several
reasons. Figure 1 plots various observables as a function of
redshift, using the WMAP cosmology (Spergel et al. 2003) and
equations in Carroll, Press, \& Turner (1992). Beyond $z\sim1$,
galaxies are only a few Gy old (top left) and a normal size bar
has only rotated around only a few times. Thus the bar will be
dynamically young and likely to appear different, perhaps more
irregular or round. Also, the bar age is likely to be younger than
the gas consumption time, which is typically $\sim10$ orbit times
(e.g., Kennicutt 1998), so the bars will still be gas-rich, like
some late-type bars today. This makes them irregular also because
of dust obscuration and star formation.

Second (Fig. 1 upper right), between $z\sim0.6$ and 4.6, objects
smaller than 2 kpc in physical size, such as a bar thickness, will
be smaller than 10 pixels on the HST ACS camera. This makes it
difficult to resolve internal structure. Bars were not easily seen
in the Hubble Deep Fields North and South (e.g., Abraham et al.
1999) because of the factor of 2 lower angular resolution of the
WFPC2 camera. Only the largest bars could be seen in the HDF
images (Sheth et al. 2003).

Third (lower left), beyond $z\sim3$ the redshift range
corresponding to the duration of bar formation exceeds $\Delta
z=1$, which means that the bar formation process gets stretched
out over a wide range of z. This is good in the sense that
different stages in bar formation may be seen distinctly, but it
is bad in the sense that bars beyond $z\sim3$ will be incompletely
formed and perhaps unrecognizable.

Fourth (lower right), beyond $z\sim4$ disks with the ``standard''
rest-B band surface brightness of today's Freeman-disk, 21.6 mag
arcsec$^{-2}$, are almost too faint to see.  The $2-\sigma$ noise
limiting surface brightness of the Hubble Ultra Deep Field is
$\sim26$ mag arcsec$^{-2}$ at I band, which is the deepest
passband (Elmegreen et al. 2005a).  At $z=4$, a Freeman disk will
have a central surface brightness of $\sim28$ mag arcsec$^{-2}$,
which requires averaging over areas at least as large as $\sim40$
square pixels to observe at the $2-\sigma$ limit. Thus only large
bright regions can be seen. Fortunately the intrinsic surface
brightnesses of disks are higher at high redshift because of
enhanced star formation, unless the extinction is also high.
Whether this star formation is enough to reveal bars at this early
epoch is unknown. Surface brightness limitations have already
diminished the number of spiral galaxies that can be seen in the
UDF (Elmegreen et al. 2005a).

The relatively late formation of disk galaxies compared to
spheroids aids somewhat in the study of bar formation because it
means that bars in disks form late too, when they can be more
easily observed. Nevertheless, there are still considerable
problems recognizing and observing young bars.  The apparent loss
of bars beyond $z\sim1$ could be partly from these selection
effects affects.

\subsection{The Morphology of Bars out to $z\sim1$}

Most bars look different at $z=1$ than they do today. High-z bars
are typically blue and often clumpy, like giant star-forming
regions (Elmegreen, Elmegreen, \& Hirst 2004). Sometimes they are
off-center. Perhaps this irregularity is not surprising since
young disks should still be gas-rich. However, the observation
implies something new about bar formation: that it could occur in
gas-rich or pure-gas disks with gas dissipation playing an
important role. The bar formation models and simulations existing
today are all in pure stellar or dominantly stellar systems, where
orbital motions and resonances are important.  In a gas rich disk,
bar formation is more like a dissipative instability. Recent
simulations of bar formation in a pure-gas disk are in Kaufmann et
al. (2004).

\section{Clumpy Structure in High-z Galaxies}

Most galaxies in the Hubble Deep fields and in the Ultra Deep
Field are very clumpy. Even elliptical galaxies are clumpy in
their cores (Elmegreen, Elmegreen, \& Ferguson 2005).  A
compendium of clumpy structures is shown in Figure 2.

\begin{figure}[ht]
\vskip.2in
\centerline{\includegraphics[width=5in]{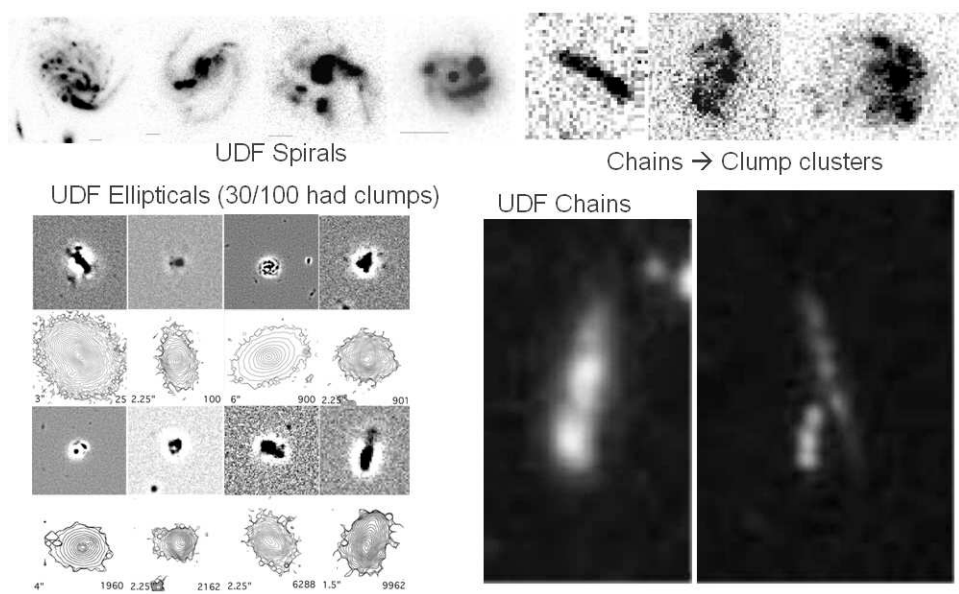}}
\caption{Mosaic of galaxy types from the HST UDF showing extremely
clumpy structure. In the lower left, elliptical galaxies are shown
with contours to illustrate the overall extents and with unsharp
masks at the same scale to highlight the clumps.}
\end{figure}

The clumps in galaxies in the UDF contain up to $10^9$ M$_\odot$
and are typically bluer than the rest of the disk or elliptical
galaxy. Thus they are probably young and still forming stars. They
are distinct from spiral galaxy bulges, which are redder and
always centrally placed.

An examination of the colors and magnitudes of clumps in 10 highly
clumpy galaxies, called ``clump clusters,'' suggests their ages
are several $\times10^8$ years (Elmegreen \& Elmegreen 2005). This
is $\sim10$ times longer than the dynamical times of the clumps,
meaning they are gravitationally bound like star clusters, but it
is not so young that they may be interpreted as currently
star-bursting. Moreover, the clumps have internal densities that
are only a factor of $\sim3-10$ above the local tidal densities in
their disks, which means they should disperse relatively soon
because of tidal forces. This is probably the process of disk
formation; i.e., through the dissolution of giant clumps (see also
Elmegreen, et al. 2005b).

The origin of the clumps is unknown at the present time. They
could be formed in the disks of spirals and clump cluster galaxies
or in the gaseous components of elliptical galaxies as a result of
gravitational instabilities (Noguchi 1999; Immeli et al. 2004a,b).
They could also come in as dwarf-like galaxies from the
surrounding field. Similar but isolated clumps are actually seen
in these fields (Elmegreen \& Elmegreen 2005; Elmegreen, et al.
2005a).

\section{Summary of Evolutionary Effects over a Hubble Time}
\label{sect-evol}

Evolutionary effects may be divided into two types, non-secular,
which result from one-time events, and secular, which result from
a progressive slow change. Non-secular events include
interactions, major mergers, and gas stripping during motion
through a galaxy cluster.

Strong interactions form bars (Noguchi 1988; Gerin, Combes, \&
Athanassoula 1990; Sundin \& Sundelius 1991; Berentzen et al.
2004). Even binary companions promote bar formation and may shift
the galaxy to an earlier Hubble type because of torqued inflow
(Elmegreen, Elmegreen, \& Bellin 1990). Bars are also more common
in perturbed galaxies (Varela et al. 2004).  Major mergers form
ellipticals (Toomre \& Toomre 1972; Barnes \& Hernquist 1992;
Schweizer 1998), and in the process, they form at least the
youngest halo globular clusters (Ashman \& Zepf 1992), and may
produce counter-rotating cores (Kannappan \& Fabricant 2001).

Minor mergers are non-secular events although they may occur more
or less continuously. They thicken spiral disks (Walker et al.
1996; Schwarzkopf \& Dettmar 2000ab; Bertschik \& Burkert 2002;
Gilmore, Wyse, \& Norris 2002) and may also lead to
counter-rotating cores. Interactions may form dwarf galaxies in
tidal tails (Zwicky 1959; Barnes \& Hernquist 1992; Elmegreen,
Kaufman, \& Thomasson 1993; Duc, Bournaud, \& Masset 2004).

Cluster ram pressure strips gas (Gunn \& Gott 1972; Warmels 1988;
Hoffman et al. 1988; Cayatte 1990; Crowl et al. 2005) and may
promote the formation of S0 types (Larson, Tinsley, \& Caldwell
1980).

Strong interactions also form ring galaxies (Theys \& Spiegel
1976; Lynds \& Toomre 1976), polar ring galaxies (e.g., Cox \&
Sparke 2004), disk warps, tidal tails, and so on (Toomre \& Toomre
1972).

Secular interactions include accretion in the form of residual
hierarchical build-up from outside the galaxy, continuous
accretion of dwarfs (e.g., Lewis et al. 2004), galaxy harassment
(e.g., Moore et al. 1996), accretion of gas (e.g., Bournaud et al.
2005), globular cluster accretion from dwarfs (e.g., Bellazzini,
Ibata, \& Ferraro 2004), halo streams (e.g., Ferguson et al.
2005), and so on.

Secular changes are also driven by internal processes, such as bar
and spiral torques, which lead to inner disk accretion and outer
disk spreading (see review in Pfenniger 2000). Gas accretion at
resonances makes star-forming rings (see review in Buta \& Combes
1996). Scattering and vertical resonances thicken disks. Bars
dissolve into lenses and (maybe) bulges, as discussed above. And
of course, gas converts into stars, while gas viscosity
redistributes the disk gas.

Secular changes among galaxies with different average densities
cause early Hubble types to age more quickly than late Hubble
types. Most evolutionary processes are faster at higher density.
As time goes by, the transition between early and late types
moves, converting little-evolved galaxies, which still have a lot
of gas and star formation, to highly evolved galaxies, which are
inactive and fading.  As galaxies progress from late to early
types, they become more centrally concentrated, have less gas,
less star formation, hotter disks, redder populations, and more
metals. Galaxy mass and size increase over time by hierarchical
buildup at $z>1$, and they probably still increase over time today
for late-type spirals as a result of continued accretion and
coalescence.  Still, the average mass and size of a galaxy that is
actively forming stars tends to decrease with time
(``downsizing'').  This is because the most massive galaxies tend
also to be the densest, and they completed their star formation
first. Also with time, galaxies become more and more isolated,
which means that flocculent spiral types and perhaps low surface
brightness disks appear relatively late.

\begin{acknowledgments}
Support from the National Science Foundation grant AST-0205097 is
gratefully acknowledged.
\end{acknowledgments}

\end{document}